\documentclass[10pt,journal,final,twocolumn]{IEEEtran}
\usepackage{amssymb}
\usepackage{amsmath}
\usepackage{epsfig}
\usepackage{algorithm}
\usepackage{graphicx}
\usepackage{subfigure}
\usepackage{bm}
\usepackage{amsthm}
\usepackage{multirow}

\newtheorem{myDef}{Definition}
\newtheorem{myTheo}{Theorem}
\newtheorem{myProp}{Proposition}
\newtheorem{myProf}{Proof}
\newtheorem{myCase}{Case}
\newtheorem{myRmk}{Remark}
\newtheorem{myProb}{Problem}

\allowdisplaybreaks

\begin{document}

\title{Towards Cooperation  by Carrier Aggregation in Heterogeneous Networks: A Hierarchical Game Approach}
\author{Pu~Yuan, Yong~Xiao, \IEEEmembership{Senior Member, IEEE}, Guoan Bi, \IEEEmembership{Senior Member}, IEEE and Liren Zhang, \IEEEmembership{Senior Member}

\thanks{P. Yuan and G. Bi are with School of Electrical and Electronic Engineering, Nanyang Technological University, Singapore (e-mails: pyuan2@ntu.edu.sg and EGBI@ntu.edu.sg).

Y. Xiao is with the Electrical and Computer Engineering at the University of Houston, TX, USA (e-mails: xyong.2012@gmail.com).

L. Zhang is with the College of Information Technology, UAE University (Email: lzhang@uaeu.ac.ae).
}
}
\maketitle

\begin{abstract}
This paper studies the resource allocation problem for a heterogeneous network (HetNet) in which the spectrum owned by a macro-cell
operator (MCO) can be shared by both unlicensed users (UUs) and licensed users (LUs). %To protect %the transmission of LUs, %the MCO imposes an
%interference power constraint on the transmit powers of UUs. Motivated by the observation that the spectrum utilization efficiency can be
%further improved by allowing multiple UUs to access the same sub-band, 
We formulate a novel hierarchical game theoretic framework to jointly
optimize the transmit powers and sub-band allocations of the UUs as well as the pricing strategies of the MCO. 
In our framework, an
overlapping coalition formation (OCF) game has been introduced to model the cooperative behaviors of the UUs. We then integrate this OCF game into a Stackelberg
game-based hierarchical framework. We prove that the core of our proposed OCF game is non-empty and introduce an optimal sub-band
allocation scheme for UUs. A simple distributed algorithm is proposed for UUs to autonomously form optimal coalition formation structure.
The Stackelberg Equilibrium (SE) of the proposed hierarchical game is derived and its uniqueness and optimality have been proved. A distributed joint optimization algorithm is also proposed to approach the SE of the game with limited information exchanges between the MCO and
the UU.
\end{abstract}

\section{Introduction}
%\subsection{Motivation}
A HetNet is a multiple tier network consisting of co-located macro-cells, micro-cells and femto-cells. It has been included in Long Term Evolution Advanced (LTE-A) standard as a part of the 
next generation mobile network technology. 
One of the motivations driving the development of HetNets is its potential to improve the spectrum utilization efficiency by reusing the existing frequency bands. 
%network coverage; support more subscribers and data services by reusing the existing frequency. 
%With rapid growth of the popularity of mobile devices, there is an urgent need for mobile networks to provide a large capacity. 
Due to the scarcity of radio resources, it is important to find an efficient method to improve the network capacity with the limited radio resources. %Hence, apply spectrum sharing in different tier of the HetNet has been considered.

%On the other hand, recently the research of applying game theory on interference control in spectrum-sharing based femto-cell networks drives our attention. 
The femto-cell is introduced to improve coverage of the cellular network as well as quality of service (QoS) of indoor mobile subscribers. The femto-cell which is controlled by a lower power BS  covers a small area and provides radio link to its own subscribers. As the deployment of the femto-cells is made by the consumers, centralized control is generally difficult to achieve. 
Game theory provides useful tools to study distributed optimization problems for multi-user network systems. Various game theoretical models have been proposed to distributedly optimize the spectrum sharing between femto-cells and existing cellular network
infrastructure \cite{kang2012price}, \cite{chandrasekhar2009power}. 
In \cite{kang2012price}, the authors modeled the distributed interference control problem as a non-cooperative game and discussed the impact of  different pricing schemes on the performance of the spectrum sharing network. 
By using Stackelberg game model, a pricing based approach to handle the interference control problem was proposed in \cite{alpcan2002cdma}, where a sub-band pricing scheme is introduced to regulate the received power at the BS for the code division multiple access (CDMA) communication system. 
However their assumption that all UUs can only access one communication channel may not always hold in practical scenarios. 

%\subsection{Game Model}
In this paper, we consider a special HetNets in which the spectrum licensed to an MCO  can be shared by mulitple co-located femto-cell base stations (BS). 
The femto-cell BSs try to make the best use of the spectrum offered by the MCO. 
The users subscribed to the service of the MCO are regarded as the LUs who have the priority to access the
resources of the MCO. The users subscribed to the femto-cell service are UUs and can only share the sub-bands owned by the MCO under the condition that the resulting interference to the LUs should be maintained below a tolerable level.
The sub-band allocation of each UU is controlled by its corresponding femto-cell BS. 
In the latest LTE-A system, the CA is introduced to support wide-band high-speed transmission by allowing multiple users to aggregate their radio resources together \cite{yuan2010carrier} \cite{xiao2013dynamic} \cite{xiao2012spatial}. Hence, we assume that each femto-cell BS can allocate multiple contiguous or non-contiguous sub-bands for each of its UUs and enabling the carrier aggregation. We also assume each sub-band can be accessed by multiple users at the same time. 
We formulate the sub-band allocation problem as an overlapping coalition formation game (OCF-game). In this game, a coalition is formed by the UUs who access the same sub-band. Since each UU can access multiple sub-bands, two or more coalitions may 
contain the same UU. In other words, the coalitions formed by UUs can be overlapped. The performance of each UU not only depends on
the sub-band allocation scheme but also its transmit power used to send signals in different coalitions.

\begin{figure}
\centering
        \includegraphics[width=0.35\textwidth]{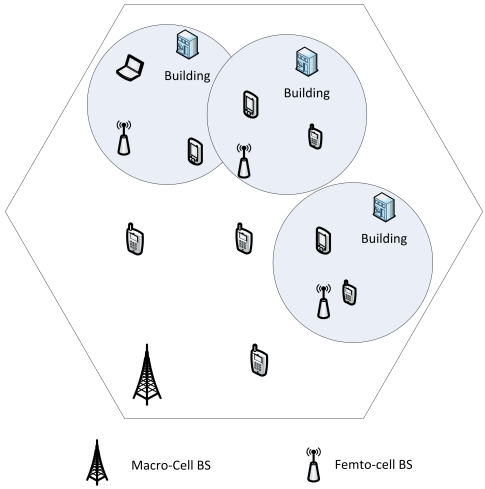}
        \caption{A spectrum sharing multi-tier HetNet in which the spectrum is owned by the macro-cell and shared with other tiers.}
    \label{fig:fig0}
\end{figure}

We integrate the OCF-game into a hierarchical game framework to investigate the interaction between the MCO and the UU. To the best of  our knowledge, this is the first work which integrates the OCF-game model into an hierarchical game framework to the analyze of HetNets. 
%The cooperation among the players in the same coalition is achieved by forming a virtual MIMO channel to transmit and receive signals cooperatively. 
%Through cooperation rather than competition, the UUs in the same sub-band achieve a better pay-off sum. Therefore, 
%The most important issue in the proposed OCF-game is to find an optimal coalition structure to maximize the pay-off sum of the femto-cell network. 

%\subsection{Main Contributions}
It is knowing that allowing overlapping among multiple coalitions will significantly increase the complexity of the game problem. Specifically, finding a stable coalition formation structure of an overlapping coalition formation game is notoriously difficult and generally requires to exhaustively search all the possible structures. In this paper,
we however show that it is possible to find the core of our proposed game under this spectrum-sharing-based HetNet without exhaustive searching.  The main contributions of this paper are summarized as follows: 
\begin{itemize}
\item[1)] A spectrum-sharing based HetNet is considered in which the  femto-cell BS can aggregate multiple sub-bands of the MCO and allocate  aggregated sub-bands to support high-speed wide-band data
transmission for each UU. This is different from our previous work \cite{xiao2010distributed} \cite{xiao2012hierarchical} where each UU is assumed to only access one sub-band. 
\item[2)] The OCF-game model is applied to study the scenario that the cooperative UUs can dedicate their power resources to multiple sub-bands.  
\item[3)] The non-emptiness of the core of our proposed OCF-game is proved, which makes the optimization of the coalition formation structure possible. 
\item[4)] A hierarchical game framework is established to study the joint optimization of transmit power and sub-band  allocation of UUs as well as the pricing strategy of the MCO. 
\item[5)] Numerical results are presented to analyze the impact of power constraint, number of users, number of available sub-bands on the performance of HetNets. 
\end{itemize}
The remaining of the paper is organized as follows. Section II introduces the related works. Section III introduces the system setup. Section
IV presents the problem formulation. Section V gives the game theoretic analysis 
and Section VI presents the distributed algorithm and Section VII shows the numerical results. 
Section VIII concludes the paper. 

\section{Related Works}
An important problem in a spectrum-sharing based network is how to give sufficient protection to the LUs of the MCO. The interference power constraint \cite{haykin2005cognitive} is usually applied to regulate the spectrum sharing between UUs and LUs. In this case the Stackelberg game can be a useful tool to model the interaction between 
the MCO and the UUs. In \cite{kang2012price} and \cite{alpcan2002cdma}, the MCO acts as the leader and has the priority to set a price to access the radio resource, and the UUs act as followers and will decide their best transmit powers based on the prices. These works show the usefulness of using Stackelberg game model in solving interference control problem with hierarchical structure, which motivated us to include the Stackelberg game into our hierarchical game framework. However the assumption that all UUs can only access one communication channel with flat-fading may not always hold in practical scenarios. In our previous work \cite{xiao2010distributed}, we focus on the case that the spectrum owned by the MCO is divided into sub-bands, while these sub-bands can be shared with the UUs. 
A non-cooperative game model to 
enable the UU to sequentially join the sub-bands while the
interference to the MCO is controlled by a pricing mechanism. The limitation of this solution is that 
the sub-band and UUs can only be one-to-one paired so frequency reuse among UUs is not considered.

The carrier aggregation (CA) is proposed to support high  peak data rate in LTE-A standard \cite{ghosh2010lte}. 
The CA technique \cite{xiao2013dynamic} \cite{xiao2012spatial}   is the process of aggregating different blocks of under-utilized spectrum into larger transmission bandwidths to support high data rate.
 The technical challenges in the implementing CA have been discussed in \cite{yuan2010carrier}. Motivated from the concept of CA, we consider the scenario that the UUs can share the sub-bands with selected UUs. 
In \cite{yuan2012price}, a similar system setup has been considered, and a heuristic algorithm is reported to achieve the Nash equilibrium of the proposed game. This work, however, only considers non-cooperative competition between UUs. In this paper, we propose a general hierarchical game theoretic framework that allows cooperation among
UUs in a distributed fashion.

The game theory based resource allocation has also been used to study  the coordination of the BSs on sub-carrier selection and interference management \cite{yang2010cognitive}, \cite{attar2011interference}. In \cite{yang2010cognitive}, the authors proposed a BS cooperation policy for nearby BSs to pick up suitable component carriers to perform CA, therefore the inter-cell interference is mitigated. In \cite{attar2011interference}, the authors analyzed the coexistence problem of macro-cell BS (primary user) and femto-cell BS (secondary user) from a cognitive radio point of view. A series of techniques, such as adaptive power transmission, non-cooperative and coalitional game, is introduced to give the solution to the interference management. However, in this paper we consider the coordination between the UUs rather than the BSs, which is a challenging task as the mobility and more population of the UUs than the BSs.

%To achieve the potential benefit gained from the cooperation
%in the mobile network, the coalitional game was introduced to
%investigate the behaviors and interactions among the nodes in
%the wireless communication systems \cite{saad2009coalitional}. In this review article, the authors
%introduced three kinds of coalitional game: the coalition game, the coalition formation game, and the graph coalition game. They pointed out  the potential of these games in modeling the wireless communication problems. This motivated us to consider coalitional game to achieve cooperatiion in tour hierarchical game framework. 

In \cite{han2009coalition} the authors studied the cooperation problem between cellular subscribers located at the middle and boundary of each cell. They found that carefully constructing pair-wise coalitions between middle and boundary nodes by allowing the middle nodes to relay the packets of boundary nodes can significantly improve the overall performance of the network. 
In \cite{la2004game}, the rate allocation problem for Gaussian multiple access
channels was investigated. %The authors enabled the users to form coalitions based on some rate split policy that would compete for the channels to maximize profit. 
The authors proved that it was possible to find an unique allocation, which always lies in the core of the game. 
In \cite{xiao2012hierarchical} the authors studied the cooperative behavior of secondary users in a two-tier spectrum sharing cognitive network where both the Stackelberg game and non-overlapping coalition formation game were combined
to build a hierarchical game framework. A joint solution was given to the sub-band allocation and interference control problem. 
%Together with the Stackelberg game between the MCO and the UUs, the authors provide a hierarchical game framework towards the solution to jointly optimize the resource allocation problem in cognitive networks. 
Although the coaltional game has been widely used to study the problems in wireless communications, most of the existing works only allow users to form disjoint coalitions. 
%In other words, denoting $\mathcal{C}_j$ as a coalition, then $\mathcal{C}_1$ and $\mathcal{C}_2$ are disjoint coalitions if $\mathcal{C}_1 \cap \mathcal{C}_2 = \emptyset$. In contrast, $\mathcal{C}_1$ and $\mathcal{C}_2$ are overlapping coalitions if $\mathcal{C}_1 \cap \mathcal{C}_2 \neq \emptyset$. 
In practical communication systems, allowing overlapping of coalitions can further improve the performance \cite{xiao2015bayesian}. 
%For example, one user $D_k$ forms coalition with $D_j$ to cooperatively transmit in sub-band $m$. If $D_k$ still has spare power, it may cooperative with $D_i$ on the sub-band $l$ to support more data rate. 
%For instance, one user $S_k$ may access multiple sub-bands and cooperate with other users in each sub-bands, hence the coalition formed in each sub-band overlapped with a common member $S_k$. 
For example, one mobile subscriber may cooperatively transmit in two different sub-bands with two different subscribers. 
However, so far only limited works have been reported to apply the overlapping coalitional game to analyze cellular networking systems. In \cite{zhang2014coalitional}, the authors studied how small cell BSs coordinate with each other to achieve efficient transmission. By allowing the femto-cells to form overlapping coalitions  
%i.e., to coordinate with different groups of other small cells, 
to jointly schedule the transmission of their subscribers, they found that the performance of mobile nodes near the cell edge was improved. 
The key difference of the proposed work comparing to the ones mentioned above is that, we adopt a new OCF-game model which enables each player to join multiple coalitions. 

\section{System Setup}
Consider an orthogonal frequency-division multiple access (OFDMA)
based two-tier network where the spectrum owned by an MCO is divided into $M$ sub-bands each of which can be accessed by multiple UUs controlled by the femto-cell BSs as illustrated in Fig. \ref{fig:fig0}. 
We denote the set of sub-bands as $\mathcal{B}$ and the set of femto-cell BSs as $\mathcal{K}$. 
Here the concept of \textit{underlay} borrowed from the cognitive radio which means that each secondary user (i.e., UU) is allowed to access the spectrum of primary users (i.e., LUs) which can tolerate limited interference from the UUs \cite{zhao2007survey}. 
In this paper, we consider frequency selective fading, i.e., channel fading in different sub-bands is interdependent. We assume the channel state 
is time-invariant and can be regarded as an constant within the duration of each time slot. 
%The additive noise in each sub-band is assumed to be white Gaussian. 
We further assume that the mobile devices are equipped with multiple antennas and hence can transmit over multiple sub-bands at the same time. Furthermore, multiple UUs are allowed to share the same sub-band with each LU. 
We perform the system analysis using numerical calculations and the simulation is running on Matlab platform.

%As the UUs can access multiple sub-bands, we denote $\mathcal{X}_{S_k} = [x^1_{S_k}, x^2_{S_k}, ...,x^n_{S_k}]$ as the indicator vector, where $x^m_{S_k}\in \{0,1\}$ indicates whether the sub-band $m$ is utilized by $S_k$. Consider that the UU experiences frequency selective fading and different channel fading via different sub-bands,  we denote $h^m_{k}$ as the channel gain between a source $S_k$ to the MBS receiver via sub-band $m$. We denote $\bm{p}_{S_k}=[p^1_{S_k}, ..., p^M_{S_k}]$ be the power allocation vector of $S_k$.
Each femto-cell BS can apply multiple sub-bands to support services for UUs, i.e., each sub-band can be accessed by the UUs from more than one femto-cell BS. We assume that in each time slot there is only one active UU
$S_k$ connected with femto-cell BS $k$. Let $h^m_{k}$ be the channel gain between $S_k$ and the macro-cell BS receiver in sub-band $m$, and $g^m_{kj}$ be the channel gain between $S_k$ and $j$th femto-cell BS. Let $\bm{p}_{S_k}=[p^1_{S_k}, ..., p^M_{S_k}]$ be the power allocation vector of UUs, where $p^m_{S_k}=0$ implies that sub-band $m$ is not used by $S_k$. Table \ref{Tab:tab1} lists the notations and symbols used in this paper . 
\begin{table}[!hbp]
\centering
\caption{The Notations}
\label{Tab:tab1}
\begin{tabular}{cl}
\hline
$\pi_{S_k}(\bm{p}_{S_k}, \bm{\mu})$ & {pay-off function of UU $S_k$} \\
\hline
$\bm{v}(\bm{p}^m, \mu^m)$ & value function of partial coalition $m$ \\
\hline
%$\bm{p}^m$ & a partial coalition $m$ \\
%\hline
%$\bm{p}_{S_k}$ & power allocation vector of UU $S_k$ \\
%\hline
$\bm{l}_{S_k}$ & sub-band allocation vector of UU $S_k$ \\
\hline
$\bm{\mu}$ & interference price vector \\
\hline
$h^m_{S_k}$ & channel gain from UU $S_k$ to macro-cell BS in \\
& sub-band $m$ \\
\hline
$p^m_{S_k}$ & transmit power of UU $S_k$ in sub-band $m$ \\
\hline
$g^m_{i,j}$ & the ratio of the channel gain between UU $i$ and BS $j$ \\ 
 & to the interference power at $k$ in sub-band $m$ \\
\hline
$\lambda^m_{S_k}$ & the pay-off division factor for UU $S_k$ in sub-band $m$ \\
\hline
$\bm{P}$ & the power allocation matrix of all UUs \\
\hline
\end{tabular}
\end{table}
Multiple femto-cell BSs can apply for the same sub-band at the same time. We denote the set of all UUs as $\mathcal{S}$. We denote the set of UUs utilizing the same sub-band $m$ as $\mathcal{L}_m$, i.e., $\mathcal{L}_m = \{S_k: p^m_{S_k} > 0, \forall S_k \in \mathcal{S}$.  $\mathcal{L}_m = \emptyset$ means no UU uses sub-band $m$, $\mathcal{L}_m = \{S_k\}$ means sub-band $m$ is exclusively 
occupied by  femto-cell BS $k$, and $|\mathcal{L}_m| \geq 2$ means sub-band $m$ has been shared by two or more femto-cell BSs. 

Different from the previous works which consider the distributed spectrum-sharing scheme \cite{yuan2012price}, UUs can cooperatively transmit the signal with co-channel peers to further improve their pay-offs. 
%Since the femto-cells may be deployed in an area without the coverage of the macro-cell, they are given the chance to aggregate their spectrum. 
In this paper, we follow the same line as \cite{mathur2008coalitions} and assume that UUs from different femto-cells sharing the same sub-band $m$ can cooperate by forming a virtual $|\mathcal{L}_m|$-input $|\mathcal{L}_m|$-output MIMO channel \cite{mathur2008coalitions}. 

%The rate achieved by the virtual MIMO channel is shown to be 
%the upper bound of the multiple access channel \cite{telatar1999capacity}.
%More specifically, at the beginning of each time slot, a portion of time is assigned for the UUs to obtain the channel state information. When UUs in the same coalition form a virtual MIMO channel, $S_k$ needs to estimate the channel gains of all the femto-cell BSs in the same coalition. We will discuss the detailed coordination protocol and information exchange among SUs in Section IV.  

In this paper, we consider the following two power constraints: 
\begin{itemize}
\item[-] Interference power constraint in each sub-band $m$,
\begin{align}
\sum_{k=1}^{K}{p^m_{S_k}h^m_{S_k}} \leq \overline{Q},
\label{co:co1}
\end{align}
where the maximum tolerable interference $\overline{Q}$ is determined by the macro-cell BS to protect the LUs.
\item[-] The transmit power cap of the mobile devices,
\begin{align}
\sum_{m=1}^{M}{p^m_{S_k}} \leq \overline{p},
\label{co:co2}
\end{align}
where $p^m_{S_k}$ is the transmit power of $S_k$ on sub-band $m$
and $\overline{p}$ is the the total amount of power that can be used by each UU $S_k$ to transmit signals. The value of $\bar p$ depends on the physical limits of the hardware as well as the battery life. %Similar setup considering both the total power and per-band  power constraints is investigated in \cite{gao2010cooperative}. 
\end{itemize}

\textit{Remark}: These two power constraints together limit the number of UUs accessing each sub-band. 
%How they jointly affect the sub-band and power allocation of the UUs depends on the particular network realization. 
For example, if $\overline{p}$ and $h^m_{S_k}$ are large, UU $S_k$ may cause interference that is close to $\overline{Q}$ so that it will be the only active UU in sub-band $m$. 
If $\overline{p}$ and $h^m_{S_k}$ are small, multiple UUs can simultaneously access the same sub-band, and the accumulated interference is still below $\overline{Q}$. The number of sub-bands used by an individual UU is affected by the power cap given in (\ref{co:co2}), but the total number of the active UUs in each sub-band is limited by the maximum tolerable interference level constraint in (\ref{co:co1}). 

The interference power constraint reflects the fact that the randomly 
distributed UUs usually give different levels of interference to each macro-cell BS. 
Due to the frequency selective fading, the interferences from the same UU are generally different in different sub-bands. 
Hence the UUs are preferred to transmit in those frequency bands with weak channel gains between the UUs and the macro-cell BS. 

An important problem is how UUs can distributedly form different coalitions to improve their pay-offs. We formulate an overlapping coalition formation game to study this problem. In this game, UUs can behave cooperatively to coordinate their actions. Hence the coalition formation game focuses on solving the following two questions: Q1) how the coalition members coordinate with each other, and Q2) how a coalition formation structure can be established among UUs. 
%\subsection{The Pay-off of the UU }

To answer the first question,  the virtual MIMO technique is used as the cooperation scheme among the UUs in the same coalition for  two reasons: 1) it is shown to achieve the 
upper-bound of the rate for a multiple access channel \cite{telatar1999capacity}, 2) it is shown to satisfy the proportional fairness \cite{xiao2012hierarchical}. More specifically, the UUs in the same sub-band $m$ form a coalition and cooperate with each other to transmit and receive signal. Using the virtual MIMO technique, we can convert the communication within one coalition into a virtual $\mathcal{L}_m$-input $\mathcal{L}_m$-output channel, which follows the same line as \cite{telatar1999capacity} and \cite{xiao2012hierarchical}. Therefore the capacity sum of all UUs in the $mth$ virtual MIMO channel is obtained as,
\begin{align}
\sum_{S_k \in \mathcal{L}_m}{r_{S_k}}
= \sum_{S_k \in \mathcal{L}_m}{\log{(1+\lambda^m_{S_k}p^m_{S_k})}},
\label{eq:sumrate}
\end{align}
where $\lambda^m_{S_k}$ is the $kth$ non-zero eigenvalue of matrix $\bm{G}^T_{\{S_k \in \mathcal{L}_m\}}\bm{G}_{\{S_k \in \mathcal{L}_m\}}$ where $\textit{\textbf{G}}_{\{S_k \in \mathcal{L}_m\}}$ is the channel gain matrix of UUs in the same sub-band. For example, if $\{S_1,...,S_n\}$ are in the same sub-band $m$, then the matrix is given by
\begin{equation}
\boldsymbol{G}_{\{S_k \in \mathcal{L}_m\}} = 
\left[
\begin{array}{cccc}
g^m_{11} & g^m_{12} & ... & g^m_{1n} \\
g^m_{21} & g^m_{22} & ... & g^m_{2n} \\
      .         &       .        & .\ \    &       .          \\
      .         &       .        & \  . \  &       .          \\
g^m_{n1} & g^m_{n2} & ... & g^m_{nn} \\
\end{array}
\right].
\label{mtransmit:mtransmit1}
\end{equation}
In the above matrix, $g^m_{jk} = \frac{g^{m\prime}_{jk}}{\sigma^m_k}$, where $g^{m\prime}_{jk}$ is the channel gain between UU $S_j$ and femto-cell BS $k$, and $\sigma^m_k$ is the received interference power at BS $k$ in sub-band $m$. 
We will give detailed analysis and propose a distributive algorithm to answer the second question in section \ref{CFGA}. 

To simplify the analysis, let us consider the uplink transmission. 
In the uplink, the receiver of macro-cell BS is interfered by the transmit signals of UUs. Therefore there is only one leader when it applies price-based interference control. However, our model can be directly extended to the downlink scenario. In the downlink case, multiple LUs act as a group of leaders which can cooperatively decide the interference price in each sub-band. 
The main objectives of this paper are to solve the following problems:
\begin{itemize}
\item[1)] {\em Power control problem}:  investigating how the MCO controls the interference power to protect the LUs by dynamically adjusting the interference price. 
\item[2)] {\em Sub-band allocation problem}: investigating how the UUs choose the sub-bands to access based on the channel information, the interference price and the action of other UUs. 
\item[3)] {\em Overlapping Coalition formation problem}: investigating how the UUs form overlapping coalitions to improve their data rate. 
\end{itemize}

We formulate a hierarchical game framework to jointly optimize the above three problems.

\section{The Hierarchical Game Formulation}
The interaction between the macro-cell BS and femto-cell BS can be modeled as a Stackelberg game. Furthermore, we also formulate a OCF-game to investigate the cooperation among the femto-cell BSs, where their UUs can form coalitions to improve the performance. 
We assume that the transmission of femto-cell and macro-cell are synchronized. 

\label{HGF}
%We first introduce the notations in table \ref{}
Our goal is to jointly solve the power control problem of the LUs and resource allocation problem of the UUs. 
Firstly, there is a trade-off between the capacity sum of the femto-cell network and QoS of the macro-cell. 
If the UUs transmit with high power, they will get high data rate but generate more interference to the macro-cell BS. 
Since sufficient protection to the LUs should be guaranteed in the first place, the MCO should  regulate the behavior of the UUs. We can  model this as a power control problem for UUs.
Secondly, given the limited spectrum and power resources, we should consider how the UUs can cooperate with each other
to allocate the sub-band and optimize their transmit power. %This results in the resource allocation problem among the UUs. 

We model a hierarchical game consisting of the two sub games. 
In the proposed game model, the MCO and femto-cell BSs are the \textit{players}. 
The way the players play the game is defined as \textit{actions}. In the proposed game model, the action of the MCO is to decide the interference prices, and the actions of the UUs are to decide which sub-bands to access and how much power should be allocated to each of these sub-bands. 
%The \textit{pay-off} is the outcome of the actions. The pay-off for the macro-cell is defined as the revenue it collected from the femto-cells, and the 
%pay-off for the femto-cells is the data rate achieved minus the cost for leasing the spectrum of the macro-cell.  

We apply the Stackelberg game to model the interaction between the MCO and the femto-cell BSs. 
In the proposed Stackelberg game, the \textit{leader} is the MCO and the corresponding LUs and the \textit{followers} are the femto-cell BSs who control the UUs. 
%\begin{myDef}
%\cite{basar1995dynamic} A Stackelberg game is a game played by a leader and followers. 
%In each round, the leader commits to a strategy based on the best responses of the followers in previous round, 
%and the followers observe the leader’s move and respond with the optimal actions, which maximize their pay-off accordingly.
%\end{myDef}
Let us follow a commonly adopted game theoretic setup \cite{kang2012price}  \cite{xiao2011simple} \cite{razaviyayn2011stackelberg} to define %the pay-off functions of the UUs, in which the \textit{benefit} is the data rate and the \textit{cost} is the payment for the interference. Both the follower and leader are selfish and try to maximize their pay-off. 
%We denote $\mu^m$ as the unit price of interference in sub-band $m$. The price in all sub-bands is denoted by vector $\bm{\mu} = [\mu^1, \mu^2, ...,\mu^M]$. 
%We adopt linear pricing scheme hence the payment from $S_k$ for interfering the macro-cell BS receiver in sub-band $m$ is proportional to the received interference power, say $\mu^m h^m_{S_k} p^m_{S_k}$. 
%The UU is benefited from transmitting on the sub-bands shared by the macro-cell, and the resulting data rate contributes to the profit term in the pay-off function, while the payment to MCO contributes to the cost term. We hence can rewrite 
the pay-off of $S_k$ as,
\begin{align}
\pi_{S_k}(\bm{p}_{S_k}, \bm{\mu}) = r_{S_k}(\bm{p}_{S_k}) - c_{S_k}(\bm{p}_{S_k}, \bm{\mu}),
\label{pay_off_UU_1}
\end{align}
where $c_{S_k}(p_{S_k, \mu})=\sum^{M}_{m=1}{\mu^m h^m_{S_k} p^m_{S_k}}$ is the cost function. Furthermore, since $S_k$ can simultaneously access multiple sub-bands, it aims to maximize the sum of the pay-offs obtained from all the active sub-bands under the constraints given in (\ref{co:co1}) and (\ref{co:co2}). 

The MCO collects the payment from all the UUs occupying the sub-bands. We define the pay-off functions of the MCO as, 
\begin{align}
\pi_{MCO}(\bm{p}_{S_k}, \bm{\mu}) =
\sum_{k=1}^{S_k}c_{S_k}(\bm{p}_{S_k}, \bm{\mu}).
\label{eq:sum_U_mo}
\end{align}

The main solution for our proposed hierarchical game is the Stackelberg equilibrium (SE) which is formally 
defined as follows: \cite{basar1995dynamic}, 
\begin{myDef}
For a fixed sub-band allocation, the pricing vector $\bm{\mu}^* = [\mu^*_1, ..., \mu^*_M]$ and the transmit power $\bm{p}^*_{S_k} = [p^{1*}_{S_k}, ..., p^{M*}_{S_k}], k= 1, ..., K$,  form a SE if the interference power constraint in (\ref{co:co1}) is satisfied, and for any $m \in \{1,..., M\}$ and $k \in \{1,..., K\}$, we have 
\begin{equation}
\mu^*_m = \arg \max_{\mu_m \ge 0} \pi_{MCO} (\bm{p}^*, \mu_m, \mu^*_{-m})
%\bm{\mu}^*_m = \arg \max_{\mu_m\geq 0} {\pi_{MCO}(\bm{p}^{*}_{S_k}, \bm{\mu})} ,
%\pi_{S_k}(\bm{p}_{S_k}^{*}, \bm{\mu}^{*}) \geq \pi_{S_k}(\bm{p}_{S_k}, \bm{\mu}^{*}) ,
\end{equation}
where $\mu_{-m}$ means all the MCO except for $m$. For any given price $\bm{\mu}$, 
%and for any given price $\bm{\mu}^*$, 
$\bm{p}^*$ is given by
\begin{equation}
\bm{p}^*_{S_k} = \arg \max_{p_{S_k} \ge 0} \pi_{S_k} (p_{S_k}, \bm{p}^*_{-S_k} ).
%\bm{p}^*_{S_k} = \arg \max_{p^{m}_{S_k}\geq 0} {\pi_{S_k}(\bm{p}_{S_k}, \bm{\mu}^{*})}, 
%\bm{p}^*_{S_k} = \arg \max_{p^{m}_{S_k}\geq 0} {\pi^m_{S_k}(\bm{p}^*_{S_k}, \bm{\mu}^{*})}, 
%\pi_{MCO}(\bm{P}^{*}, \bm{\mu}^{*}) \geq \pi_{MCO}(\bm{P}^{*}, \bm{\mu}).
\end{equation}
\end{myDef}

\begin{figure}
    %\centering
        \includegraphics[width=0.5\textwidth]{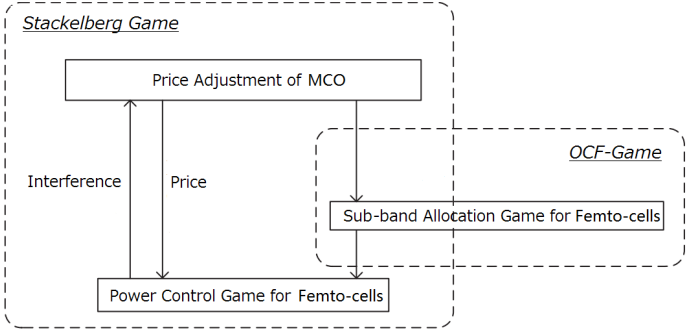}
        \caption{The hierarchical game structure.}
    \label{fig:fig1}
\end{figure}

The structure of the hierarchical game is illustrated in Fig. \ref{fig:fig1}.
The MCOs can adjust their prices to maximize the pay-off
defined in (\ref{eq:sum_U_mo}). We will show that the optimal price is specified 
by the dynamics of the interference from the CA in
each sub-band. The femto-cells BSs can cooperate and self-organize into
coalitions each of which consists of member UUs to coordinate
the transmission to improve the sum of the pay-offs. 
%On the MCO side, the price is adjusted to maximize the pay-off in (\ref{eq:sum_U_mo}). We will show that the maximization can be achieved by catching the dynamic of the aggregated interferences in each sub-band, without obtaining global information. 
%If the aggregated interference in a sub-band is higher than the interference constraint, the MCO will increase the channel price to scare away the UUs. If the interference level is low, the MCO can lower the price to encourage the UUs to transmit with more power, or attract more UUs to utilize this band to earn the benefits. Each UU calculates the optimal transmit power based on the sub-band prices. 
%Generally speaking, the transmit power of a particular UU decrease with the sub-band prices. 
%So the burden of information exchange between MCO and UU are released. The MCO only needs to measure the interference and gives feedback to the UUs with the price. 
On the femto-cells BS side, they cooperate and self-organize into coalitions, in which 
their member UUs can coordinate their transmission to improve the sum of pay-off. 
%It is assumed that the pay-off inside the coalition are transferable, therefore all member subscribers aim to maximize the sum of pay-off. 
%The details are to be provided in Section \ref{CFGA}. 

\subsection{The pay-off of UU}
\label{sec:payoff_UU}
%We start from introducing the pay-off functions of the proposed hierarchical game. 
Suppose that the overlapping coalition formation structure is fixed. Each $S_k$ having already obtained a fixed $\lambda_{S_k}$, 
%Furthermore, from (\ref{pay_off_UU_1}), (\ref{eq:sumrate}) and (\ref{eq:singlerate}) we can see that, once $\lambda^m_{S_k}$ is fixed, then maximize 
%the pay-off sum can be decomposed into a series of individual maximization problems correspond to the UUs. 
we can write the payoff of each UU $S_k$ as
\begin{equation}
\pi^m_{S_k}(p^m_{S_k,}, \mu^m, \lambda^m_{S_k}) = \log{(1+\lambda^m_{S_k}p_{S_k})}-\mu^m h^m_{S_k} p^m_{S_k}. 
\label{eq:pay_off_us_m}
\end{equation}

The optimal power allocation of $S_k$ is obtained by solving the following optimization problem, 
\begin{myProb}
\begin{align}
&\max_{\bm{p}_{S_k}} \pi_{S_k}(\bm{p}_{S_k}, \bm{p}_{-S_k}, \bm{\mu}, \bm{\lambda}_{S_k}) \nonumber \\
%& = \sum_{m=1}^{M}{r^m_{S_k} - c^m_{S_k}} \nonumber  \\
%& = \sum_{m=1}^{M}{\left(\log{(1+\lambda^m_{S_k}p_{S_k})}-\mu^m h^m_{S_k} p^m_{S_k}\right)} ,\label{eq:pay_off_UU} \\
& \mbox{S.t.} \ \ \ \sum_{m=1}^{M}{p^m_{S_k}} \leq \overline{p}. \nonumber 
%& \ \ \ \  \ \ \ p^m_{S_k}\geq 0. \nonumber 
\end{align}
\label{prob:pay_off_us}
\end{myProb}

In the proposed Stackelberg game framework, the maximum tolerable interference in (\ref{co:co1}) is omitted in Problem \ref{prob:pay_off_us} because it is included in the interference $\mu^m$ and thus is autonomously satisfied. 
Hence we only need to consider the constraint in (\ref{co:co2}). 
%Note that we can not directly apply (\ref{co:co2}) to Problem \ref{prob:pay_off_us} to obtain the power allocation by using water-filling solution because the power value does not satisfy the requirement to form a virtual MIMO channel.  

Problem \ref{prob:pay_off_us} can be directly solved by using the standard convex optimization approaches and the resulting
optimal transmit power for UU $S_k$ in sub-band $m$ is given by,
\begin{align}
p^{m\dag}_{S_k} & = \arg \max_{p^m_{S_k} \geq 0} \pi^m_{S_k}(p^m_{S_k,},\mu^m, \lambda^m_{S_k}) \\
& = \left(\frac{1}{\mu^{m} h^m_{S_k}}-\frac{1}{\lambda^m_{S_k}} \right)^{+}.
%\\m=1,2,...,M. 
\label{eq:opt_pw1}
\end{align}
%The power allocated to each UU $S_k$ in each sub-band $m$ should be given by $p^{m\dag}_{S_k}$ only. In other words, if $S_k$ decides to access sub-band $m$, it can only transmit with the power $p^{m\dag}_{S_k}$, or it will cause the failure of the coalition. 
%Then the sub-bands accessed by $S_k$ are chosen to satisfy the constraint in (\ref{co:co2}). 

We write $\bm{p}^{\dag}_{S_k} = [p^{1\dag}_{S_k}, p^{2\dag}_{S_k}, ..., p^{m\dag}_{S_k}]$. 
Due to the power cap constraint in (\ref{co:co2}), the final power allocation will fall into the following two cases: 
\begin{myCase}
$\sum_{m=1}^{M}{p^{m\dag}_{S_k}}\leq \overline{p}$. %In this case, the correlation between channels is eliminated. 
In this case, $S_k$ can access all sub-bands under the constraint defined in (\ref{co:co2}).  The power allocation of $S_k$ is decided by constraint in (\ref{co:co1}). Hence we can remove (\ref{co:co2}) and the power allocation of the UU solely depends on the sub-band prices. 
Each of the UUs tries to solve (\ref{eq:pay_off_us_m}) for the optimal power allocation and obtain $p^{m\dag}_{S_k}$ to maximize the pay-off. 
\label{case:case1}
\end{myCase}

\begin{myCase}
$\sum_{m=1}^{M}{p^{m\dag}_{S_k}}> \bar{p}$. In this case, only selected sub-bands can be accessed by the UU $S_k$. More specifically, the solution is achieved by 
searching a sub-set $\mathcal{N}_i \subset \mathcal{M}$ such that the following condition is satisfied: 
%\begin{equation}
%\left\{
%\begin{array}[c]{l}
%\sum \limits_{m \in \mathcal{N}_i}{p^{m\dag}_{S_k}} \leq \overline{p} \\
%\sum \limits_{m \in \mathcal{N}_i}{\pi(p^{m\dag}_{S_k})} \geq \sum \limits_{n \in \mathcal{N}_j, j\neq i}{\pi(p^{n\dag}_{S_k})}, 
%\end{array}
%\right.
%\end{equation}
\begin{equation}
\sum \limits_{m \in \mathcal{N}_i}{\pi(p^{m\dag}_{S_k})} \geq \sum \limits_{n \in \mathcal{N}_j, j\neq i}{\pi(p^{n\dag}_{S_k})}, 
\end{equation}
where $\{\mathcal{N}_j\}$ denotes the set of all possible sub-sets of $\mathcal{M}$ except $\mathcal{N}_i$. 
This case implies that once the price is fixed, the number of sub-bands accessed by one UU is bounded by the power cap constraint, and obviously we have $\sum \limits_{m \in \mathcal{N}_i}{p^{m\dag}_{S_k}} \leq \overline{p}$. 
\label{case:case2}
\end{myCase}

In either cases, we can obtain the optimal power allocation of $S_k$:  
\begin{align}
\bm{p}^*_{S_k} = \{p^{m*}_{S_k},  m = 1, 2, ..., M.\}, \nonumber  \\ 
p^{m*}_{S_k} = 
\left\{
\begin{array}[c]{l}
p^{m\dag}, \mbox{if } m \in \mathcal{N}_i \\ %\cap \mathcal{M}, \\
0, \mbox{otherwise.}
\end{array}
\right.
\label{eq:power_optt}
\end{align}

The corresponding sub-band allocation indicator is,
\begin{align}
\bm{l}^*_{S_k} = \{l^{m*}_{S_k},  m = 1, 2, ..., M.\}, \nonumber  \\ 
l^{m*}_{S_k} = 
\left\{
\begin{array}[c]{l}
1, \mbox{if } \bm{l}^*_{S_k} > 0, \\
0, \mbox{otherwise.}
\end{array}
\right.
\label{eq:sb_optt}
\end{align}

From the results above, it can be observed that the optimal solution of the transmit power only depends on the values of $\mu^m$ and 
$\lambda^m_{S_k}$. 
The prices are decided by the MCO through its interaction with UUs, and $\lambda^m_{S_k}$ is obtained from coalition formation structures of UUs. In rest of this section, we discuss how to obtain optimal $\mu^m$ and $\lambda^m_{S_k}$.

\subsection{The pay-off of the MCO}
\label{SGA}
The MCO can use the prices $\bm{\mu}$ charged to the UUs to control the interference in each sub-band. We will show that the MCO can maximize its pay-off by adjusting the prices based on the dynamic of the aggregated interference at the macro-cell BS receiver. Hence the proposed algorithm greatly reduces the communication overhead and makes the distributed power allocation approach possible. 
 
The revenue gained by the MCO by sharing sub-band $m$ is given by: 
\begin{align}
\pi_{MCO}(\bm{p}^m, \mu^m) =  \mu^m \sum^{K}_{k=1}{h^m_{S_k}p^m_{S_k}}. 
\end{align}

Hence the MCO tries to find the optimal sub-band price to maximize its revenue in each sub-band under the maximum tolerable interference constraint. 
\begin{myProb}
\begin{align}
&\max_{\mu^m} \  \pi_{MCO}(\bm{p}^m, \mu^m)\\
& \ \mbox{s.t. } \sum_{k=1}^{K}{p^m_{S_k}h^m_{S_k}} \leq \overline{Q}. \\
& \ \ \ \ p^m_{S_k} \geq 0. \label{eq:pay_off_mo}
\end{align}
\label{prob:pay_off_mo}
\end{myProb}

Substitute (\ref{eq:opt_pw1}) into Problem \ref{prob:pay_off_mo}, we obtain, 
\begin{myProb}
\begin{align}
&\max_{\bm{\mu}} \  \sum_{k=1}^{K}{\left(\frac{1}{h^{m}_{S_k}}-\frac{\mu^m}{\lambda^m_{S_k}} \right)^{+}}h^m_{S_k} \\
& \ \mbox{s.t. } \sum_{k=1}^{K}{\left(\frac{1}{\mu^{m} h^m_{S_k}}-\frac{1}{\lambda^m_{S_k}} \right)^{+}}h^m_{S_k} \leq \overline{Q} \label{co:co_Q2},
\end{align}
\label{prob:pay_off_mo1}
\end{myProb}

Using standard convex optimization approach to find the optimal $\mu^m$ in above problem requires the MCO to obtain global information of the UUs. Fortunately, Problem \ref{prob:pay_off_mo1} has a nice property that the objective and constraint functions both monotonically decrease with $\mu^m$. Hence if we assume the power cap constraint is satisfied, then the objective function will be maximized when the constraint in (\ref{co:co_Q2}) takes equality. 
Note that the left side of (\ref{co:co_Q2}) is the aggregated interference received by macro-cell BS in sub-band $m$. Therefore the MCO can optimize price $\mu^m$ and affect the
%Based on the analysis in subsection \ref{HGF}, we notice that the Stackelberg game model enables simple and distributed 
%pay-off optimizations. The MCO 
 aggregated interferences to the upper bound. 
% while the femto-cell BSs negotiate with each other to form overlapping coalitions and adjust the transmit power adapting to the price. 

\section{Coalition Formation Game Analysis}
\label{CFGA}
In this section, we first define the coalitional game and imputation, and then analyze the game properties to prove the existence of the core. 
\begin{myDef}
[\cite{myerson2013game}, Chapter $9$]
A coalition $\mathcal{C}$ is a non-empty sub-set of the set of all players $\mathcal{K}$, i.e., $\mathcal{C} \subseteq \mathcal{K}$. A coalition of all players is referred as the grand coalition $\mathcal{K}$. 
A coalitional game is defined as $(\mathcal{C}, v)$ where $v$ is the value function mapping a coalition structure 
$\mathcal{C}$ to a real value $v(\mathcal{C})$. A coalitional game is
said to be super-additive if for any two disjoint coalitions $\mathcal{C}_1$ and $\mathcal{C}_2$, $\mathcal{C}_1 \cap \mathcal{C}_2 = \emptyset$ and $\mathcal{C}_1, \mathcal{C}_2 \subset \mathcal{K}$, we have, 
\begin{equation}
v(\mathcal{C}_1 \cup \mathcal{C}_2) \geq v(\mathcal{C}_1) + v(\mathcal{C}_2). 
\end{equation}
Given two coalitions $\mathcal{C}_1$ and $\mathcal{C}_2$, 
we say $\mathcal{C}_1$ and $\mathcal{C}_2$ overlap if $\mathcal{C}_1 \cap \mathcal{C}_2 \neq \emptyset$. 
\end{myDef}
\begin{myDef}
A pay-off vector $\bm{\pi}$ is a division of the value 
$v(\mathcal{C})$ to all the coalition members, i.e., $\bm{\pi} = [\pi_{S_1},\cdots ,\pi_{S_K}]$. We say $\bm{\pi}$ is group rational if $\sum^{K}_{k=1}{\pi_{S_k}} = v(\mathcal{C})$ and individual rational if $\pi_{S_k} \geq v(\{S_k\}), \forall S_k \in \mathcal{C}$. We define an imputation as a pay-off vector satisfying both group and individual rationalities. 
\end{myDef}

If a coalitional game satisfies the supper-additive condition, all the players will have the incentive to form a grand coalition.
However if the supper-additive condition does not hold, then the grand coalition will not be the optimal solution for all players.
In this case, the players will try to form a stable coalition formation structure in which no player can profitably deviate from it. 
%This is different from the one defined in the coalitional game which is a set of imputations stabilizing the grand coalition. 
%In practical problems such as the proposed scenario, the grand coalition is usually not optimal. 
In the proposed OCF-game, for each possible prices of the MCO,  we focus on finding optimal coalition formation structure, for UUs to share the spectrum of the MCO. 

When overlapping is enabled among coalitions, the coalitions are no longer disjoint sub sets of the player set as defined in the non-overlapping coalitional game. In the OCF-game, the concept \textit{partial coalition} is utilized: 
\begin{myDef}
The partial coalition is defined as a vector $\bm{p}^m = (p^m_{S_1}, p^m_{S_2}, ..., p^m_{S_K})$, where 
$p^m_{S_k}$ is the fractional resource of $S_k$ dedicated to coalition $m$. Note that $p^m_{S_k}=0$ means $S_k$ is not a member of the $m$th coalition. 
A coalition structure is a collection $\bm{P} = (\bm{p}^1, ..., \bm{p}^M)$ of partial coalitions. 
\end{myDef}

\begin{myRmk}
In a non-overlapping coalition formation game, a coalition is just a subset of the player set. For a player set of size $N$, the number of coalition formation structures is given by the Bell number $B_N$, where $B_N = \sum_{k=0}^{N-1}\binom{N-1}{k}B_k$ is the number of possible coalition structures and $B_k$ is the number of ways to partition the set into $k$ items. 

For example, for a game with two players $S_1$ and $S_2$, the possible partitions can be written as  
$\{S_1, S_2\}$ or $\{\{S_1\}, \{S_2\}\}$. However, in OCF-game the concept of partial 
coalition not only specifies who joins each coalition, but also indicates how much resource each player will allocate to each coalition. 
If the resource is continuous, there are generally an infinite number of partial coalitions. %For example, for players set $\{S_1, S_2\}$, 
%the set of partial coalitions may be $\{\{0, 1\}, \{0.2, 0.3\}, \{1,1\}, \{0.5, 0\}, ...\}$. 
It means that the concept of coalition can be regarded as a special case of the partial coalition, where each player joins only one coalition with all its resource. 
\end{myRmk}

\begin{myDef}
An OCF-game is denoted by $G = (\mathcal{K}, \mathcal{M}, \bm{P}, \bm{v})$, where 
\begin{itemize}
\item[-] $\mathcal{K} = \{1, 2,..., K\}$ is the set of players which are the femto-cell BSs.  
\item[-] $\mathcal{M} = \{1, 2,..., M\}$ is the set of sub-bands. 
\item[-] $\bm{P}$ is the power allocation matrix, where the row $\bm{p}_{S_k} = (p^1_{S_k}, p^2_{S_k}, ..., p^M_{S_k})$ represents how player $S_k$ 
assign its power on different sub-bands, and the column $\bm{p}^m = (p^m_{S_1}, p^m_{S_2}, ..., p^m_{S_K})$ represents the power each player consumes for sub-band $m$. 
$\bm{p}^m = (p^m_{S_1}, p^m_{S_2}, ..., p^m_{S_K})$ also corresponds to a partial coalition. 
\item[-] $\bm{v}(\bm{C}^m): \mathbb{R}^n\longrightarrow \mathbb{R}^+$ is the value function, which represents 
the total pay-off of a partial coalition $\bm{C}^m$. %\item[-] $\pi(p^m_{S_k})$ is the pay-off of player $S_k$ on .   
\end{itemize}
\end{myDef}

\begin{myDef}
We define a game to be \textit{U-finite} if for any coalition structure that arises in this game, the number of all possible partial coalitions is bounded by $U$.  
\end{myDef}

\begin{figure}
    \centering
        \includegraphics[width=0.4\textwidth]{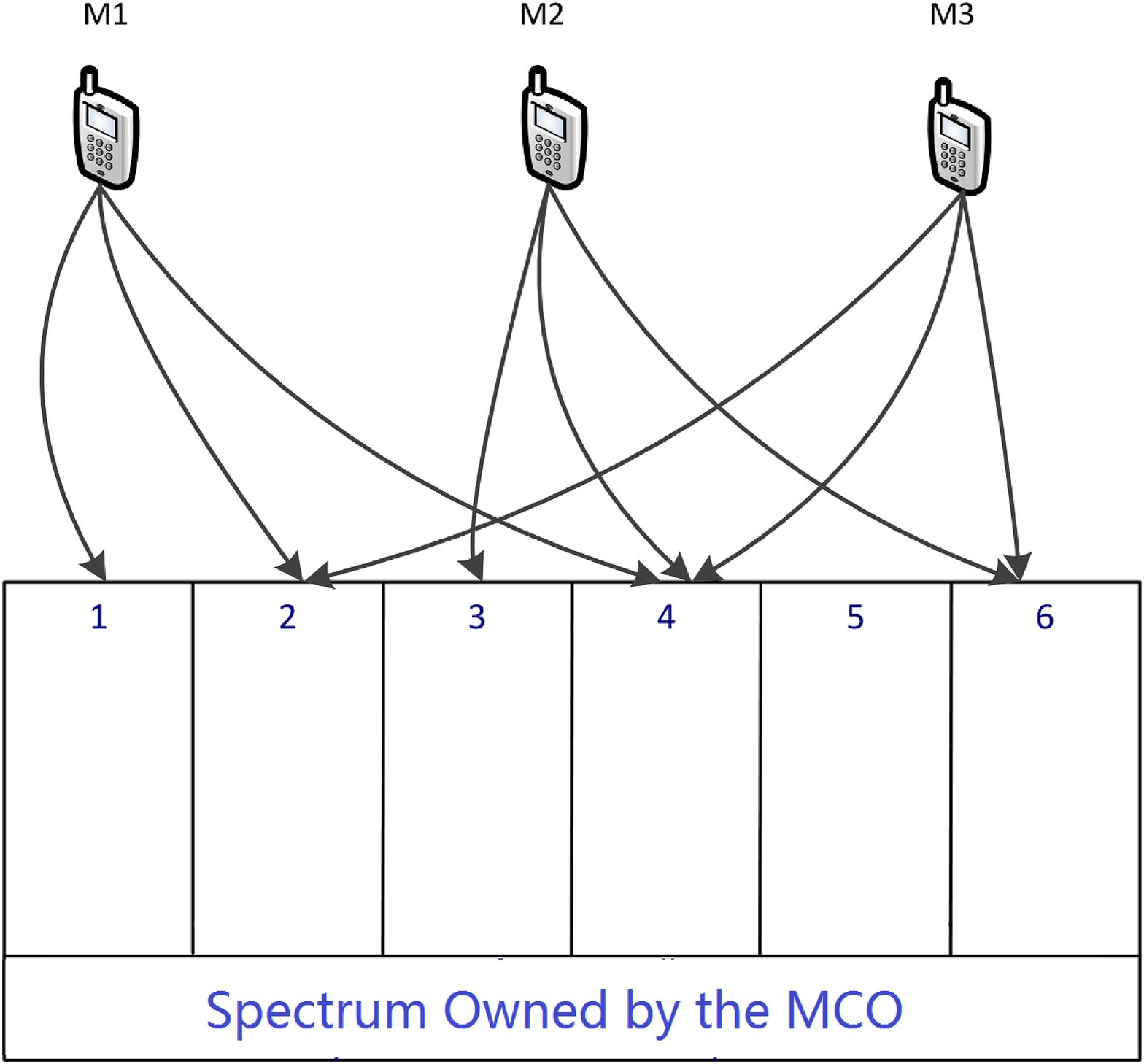}
        \caption{The illustration of the overlapping coalitions in our proposed game.}
    \label{fig:fig_OCF}
\end{figure}

%Before we go into the analysis of our model, we first give some definitions in the proposed OCF-game model. 
Fig. \ref{fig:fig_OCF} illustrates an example of the overlapping coalition formation of our model. The spectrum of the MCO is divided into six sub-bands $\{1,2,3,4,5,6\}$ which can be allocated to three mobile devices $\{M1, M2, M3\}$. A coalition is formed on the sub-band if it is accessed by two or more mobile devices. 
Each mobile device may belong to multiple coalitions, since it may access multiple sub-bands at the same time. We say the coalitions containing a common member player are overlapping. 
For example, in Fig. \ref{fig:fig_OCF}, we denote the coalition formed by the devices accessing sub-band $k$ as $\mathcal{C}_k$, Then we have,  $\mathcal{C}_1=\{M1\}$, $\mathcal{C}_2=\{M1, M3\}$, 
$\mathcal{C}_3=\{M3\}$, $\mathcal{C}_4=\{M1, M2, M3\}$, $\mathcal{C}_6=\{M2, M3\}$, $\mathcal{C}_5 = \emptyset$. 
Hence, $\mathcal{C}_1$, $\mathcal{C}_2$ and $\mathcal{C}_4$ overlap with each other since $\mathcal{C}_1 \cap \mathcal{C}_2 \cap \mathcal{C}_4 = M1$. Similarly, $\mathcal{C}_3 \cap \mathcal{C}_4 \cap \mathcal{C}_6 = M2$ and $\mathcal{C}_2 \cap \mathcal{C}_4 \cap \mathcal{C}_6 = M3$. 

The sum rate achieved by forming coalition is given by (\ref{eq:sumrate}), and the pay-off sum of UUs equals to the sum rate minus the payment to the MCO. 
Hence the value function of the partial coalition $\bm{p}^m$ is defined as the pay-off sum on sub-band $m$. Given the fixed price vector $\bm{\mu}$, the value function of the partial coalition $\bm{p}^m$ is given by, 
\begin{align}
\bm{v}(\bm{p}^m, \bm{\lambda}^m) = \sum_{S_k \in \mathcal{L}_m}{r_{S_k}} - \sum_{S_k \in \mathcal{L}_m} \mu^m h^m_{S_k}p^m_{S_k}. 
\label{eq:value_fun}
\end{align}

It is proved in \cite{xiao2012hierarchical} that the pay-off division among coalition members satisfies the proportional fairness \cite{kelly1998rate} and if the benefit allocated to each member equals to its contribution to the overall rate in sub-band $m$, i.e.,
\begin{align}
r^m_{S_k}= \log{(1+\lambda^m_{S_k}p^m_{S_k})}. 
\label{eq:singlerate}
\end{align}

The solution of the optimal power vector $p^m_{S_k}$ of $S_k$ is given by (\ref{eq:power_optt}), which is a function of $\lambda^m_{S_k}$ and $\mu^m$. Since $\bm{\mu}$ is imposed by the MCO, the UUs can optimize their pay-off sum by choosing proper sub-bands to access. Furthermore, since $\lambda^m_{S_k}$ is decided by the coalition structure, finding sub-band allocation will directly affect the payoff of each UU.  %Since the sets of UUs, the sub-bands and $\lambda^{m}_{S_k}$ are all finite discrete sets, the searching complexity is therefore limited. 

There are two types of actions of the players in an OCF-game, which are the coalitional action and the overlapping action. The former defines how the resource being allocated among the member players in one coalition, and the latter defines how resources being allocated between players in the overlapping parts of multiple coalitions.  These are the key features to differentiate the OCF-game from the non-overlapping coalition formation game. 

In the proposed system setup, the femto-cell BSs whose UUs are accessing the same sub-band form a coalition. 
The cooperation among the member players is achieved by forming a virtual MIMO channel. 
The pay-off division relies on assigning $\bm{\lambda}$ to the players, which can be considered as the contribution of  each coalition member to the sum rate. 
Since the UUs can join multiple coalitions, the proposed game becomes an OCF-game. 
The resource of a UU is the total transmit power. 
The UUs need to allocate its transmit power in each sub-band properly for  maximizing the pay-off. 
For the proposed OCF-game, we have the following definition. 
\begin{myDef}
For a set of UUs $\mathcal{S}$, a coalition structure on $\mathcal{S}$ is a finite list of vectors (partial coalitions) $\bm{P} = (\bm{p}^1, ..., \bm{p}^M)$ that satisfies (i) $\sum^{K}_{k=1}{h^m_{S_k}p^m_{S_k}} \leq \overline{Q}$, (ii) $\sup{ \bm{p}^m} \subseteq \mathcal{S}$ for all $m = 1, ..., M$, and (iii) $\sum^{M}_{m=1}{p^m_{S_k}\leq \overline{p}}$ for all $j\in \mathcal{S}$. 
\end{myDef}

The power allocation matrix also indicates the utilization status of sub-bands. The constraint $(i)$ states that the transmit power of UU in each sub-band is bounded, $(ii)$ states that the overlapping coalition is a subset of the grand coalition, and $(iii)$ states that the sum of transmit power is upper bounded. %The overlapping of the coalition is enabled. 

%Furthermore, the players are not necessarily transmit at full power, i.e., it can be the case that  
%$\sum^{M}_{m=1}{p^m_{S_k}} < \overline{p}$, since the concavity of the pay-off function in each sub-band and the maximum pay-off sum may not be achieved at maximum transmit power. Therefore the pay-off function may not have the monotone property. 
\begin{myProp}
The proposed OCF-game is $2^K\mbox{-finite}$.  
\label{prop:U_finite}
\end{myProp}

\begin{myProf}
See Appendix A. \qed
\end{myProf}
The above result suggests that it is possible to reduce the number of possible coalition formation structures into a finite set.

%This is an important result since the coalition structure is reduced to a finite set, which enables us to find the core of the proposed game. In traditional coalitional game studying the grand coalition which is a finite set of all players, the core is a set of imputations, i.e., efficient pay-off division vectors satisfying individually rationality, which stabilizes the grand coalition.  %under the super additive assumption. 
%However, many practical problems are naturally inefficient with the cooperation of all players.
We are interested in investigating a stable coalition structure which optimizes the pay-off sum.  %i.e., to find an optimal partitioning of the grand coalition.
Following the same line in \cite{chalkiadakis2008overlapping}, let us define the core of the OCF-game for the sub-bands allocation,
\begin{myDef}
For a set of players $\mathcal{I} \subseteq \mathcal{K}$, a tuple $(\bm{P}_{\mathcal{I}}, \bm{\pi}_{\mathcal{I}})$ is in the core of an OCF-game $G = (\mathcal{K}, \bm{v})$. If for any other set of player $\mathcal{J} \subseteq \mathcal{K}$, any coalition structure $\bm{P}_{\mathcal{J}}$ on $\mathcal{J}$, and any imputation $\bm{y}_\mathcal{J}$, we have $p_j(\mathcal{C}_{\mathcal{J}}, \bm{y}_{\mathcal{J}}) \leq p_i(\mathcal{C}_{\mathcal{I}}, \bm{\pi}_{\mathcal{I}})$ for some player $j\in J$.
\end{myDef}
%The imputation set $I(\mathcal{C}_{\mathcal{J}})$ corresponds to pay-off divisions among the players. In our model, it relates to $\lambda^m_{S_k}$. 

\begin{myTheo}
\cite{chalkiadakis2008overlapping} Given an OCF-game $G = (\mathcal{K}, \bm{v})$, if $\bm{v}$ is continuously bounded, monotone and U-finite for some $U \in \mathbb{N}$, then an outcome $(\mathcal{C}_{\mathcal{S}}, \bm{\pi})$ is in the core of $G$ iff $\forall S \in N$,
\begin{align}
\sum_{j\in S}p_j(\mathcal{C}_{\mathcal{S}}, \bm{\pi})\leq v^*(S),
\label{eq:eq_16}
\end{align} 
where $v^*(S)$ is the least upper bound on the value that the members of $S$ can achieve by forming the coalition.
\end{myTheo}

\begin{myProp}
The core of the proposed OCF-game is non-empty. 
\label{prop:OCFG}
\end{myProp}

\begin{myProf}:
See Appendix B. \qed
\end{myProf}

Since enabling overlapping in the coalition formation game will significantly increase the complexity of the game, the overlapping coalition structure is sometimes unstable as there may exist cycles in the game play. For example, let us consider a network system with three UUs $S_1$, $S_2$ and $S_3$, and two sub-bands $l_1$ and $l_2$. We denote $\pi_{S_j}[m|S_i]$ as the pay-off obtained by $S_j$  when it forms coalition with $S_i$ on sub-band $m$, and $\pi_{S_j}[m|\emptyset]$ is the pay-off obtained by $S_j$ when it exclusively occupies $m$. Initially, since $\pi_{S_1}[l_1|\emptyset] > \pi_{S_1}[l_2|\emptyset]$, $\pi_{S_2}[l_2|\emptyset] > \pi_{S_2}[l_1|\emptyset]$ and $\pi_{S_3}[l_2|\emptyset] > \pi_{S_3}[l_1|\emptyset]$, $S_1$ joins $l_1$, $S_2$ and $S_3$ join $l_2$. However, if we assume the following statements hold for the three UUs, 
1) $\pi_{S_1}[{l_2}|S_2] > \pi_{S_1}[l_1|S_3]$ and $\pi_{S_1}[{l_1}|S_2] > \pi_{S_1}[l_2|S_3]$, 2) $\pi_{S_2}[l_1|S_3] > \pi_{S_2}[l_2|S_2]$ and $\pi_{S_2}[l_2|S_3] > \pi_{S_2}[l_1|S_2]$, 3)$\pi_{S_3}[l_1|S_1] > \pi_{S_3}[l_2|S_2]$, $\pi_{S_3}[l_2|S_1] > \pi_{S_3}[l_1|S_2]$, then we can easily observe that the game play of the coalition formation will be stuck in a cycle. To avoid this situation, a history of the coalition structure is maintained in the proposed algorithm. If a rotation is detected, it will be removed from the coalition formation flow.

\section{Coordination Protocol Design and Distributed Algorithm}
In this section, we discuss the protocol design of the UUs' coordination   and distributed  algorithms which can reach the coalition structure in the core of the coalition formation game and the SE of the hierarchical game. 

\subsection{The Protocol Design for Coordination of UUs}
To implement the proposed algorithm into more practical systems, we consider the MAC protocol  in this section..  
We have the following assumptions: 
\begin{itemize}
\item We follow the same line as in \cite{zhao2005distributed} %\textit{[Jun Zhao, Haitao Zheng, and Guang-Hua Yang. Distributed coordination in dynamic spectrum allocation networks. In IEEE International Symposium on New Frontiers in Dynamic Spectrum Access Networks (DySPAN). Baltimore, Maryland, USA, 2005.]} 
to introduce the following distributed coordination scheme among UUs. More specifically, the UUs accessing the same sub-bands perform in-band communication with each other, both control packets and data packets are transmit in the same channel, hence there is no need for a dedicated control channel. 
\item We follow the same line as in \cite{xiao2012hierarchical} %\textit{[Y. Xiao, G. Bi, D. Niyato, and L. A. DaSilva, “A hierarchical game
%theoretic framework for cognitive radio networks,” IEEE Journal on
%Selected Areas in Communications, vol. 30, no. 10, pp. 2053–2069,
%2012.]} 
and \cite{lin2011random} %\textit{[K. Lin, S. Gollakota, and D. Katabi, “Random access heterogeneous MIMO networks,” in ACM SIGCOMM’11, 15-19 August, 2011, Toronto, Ontario, Canada.]} 
and assume that the channel gain between each UU and femtocell BS is the
same in both forward and backward directions.
\item We assume that the channel gain can be regarded as a constant within one time slot. Each time slot consists of the duration for control packets exchange and data packets transmission. 
\end{itemize} 

We introduce two control packets, request-to-send (RTS) and clear-to-send (CTS), for UUs
sharing the same sub-band to exchange their identity and establish coordination links with each other. Each control packet also contains the address information of the transmitter so the UU can identify the source of the packet. Each UU can extract the channel gain information from its received control packet.

\begin{itemize}
\item[Step 1)]The channel gain estimation and neighborhood discovering:  
\begin{itemize}
\item[a)] Firstly, the femto-cell BSs broadcast the RTS packet to all the UUs for them to estimate the channel gain. %For example, the BS $k$ sends RTS packet to UU $S_j$ over sub-band $m$, based on the $p_0$ and the receive signal strength, this channel gain $g^{m\prime}_{jk}$ is calculated by $S_j$. 

\item[b)] Each UU can then utilize the control packets for the in-band neighbor discovering and channel gain information exchange. 
For example, UU $S_j$ sends $g^{m\prime}_{jk}$ and $g^{m\prime}_{jj}$ to UU $S_k$ in sub-band $m$. Upon receiving the information sent by $S_j$, $S_k$ will then send back a CTS packet containing $g^{m\prime}_{kj}$ and $g^{m\prime}_{kk}$ to $S_j$. Hence $S_j$ knows that $S_k$ is also accessing sub-band $m$ as well as the channel gain information. 
\end{itemize}
\item[Step 2)] Coalition formation: 
\begin{itemize}
\item[a)] After the channel estimation and neighbor discovering, the UUs need to calculate and negotiate the pay-off division factor $\lambda^m_k$. Since the channel gain and neighborhood information is obtained in previous step, each of the UUs can construct $G_{S_{k}\in \mathcal{L}m}$ and subsequently calculate $\lambda^m_k$. The assignment of $\lambda^m_k$ to each UU $S_k$ could be random or follow some policies \cite{xiao2012hierarchical}. Here we assign the $\lambda^m_k$ following the rank of channel gains. Suppose the pay-off division vector $\bm{\lambda}^m$ is sorted in ascending order $[\lambda^m_1,... ,\lambda^m_K]$. The UU $S_k$ has already obtained the channel gain $g^{m\prime}_{jj}, j = 1,... ,K$ in step 1). $S_k$ sorts the channel gain in ascending order and finds the rank value $r_{S_k}$ of $g^{m\prime}_{kk}$. Then it picks the $r_{S_k}$th element in $\bm{\lambda}^m$ as its pay-off factor, i.e., $\lambda^m_k = \bm{\lambda}^m[r_{S_k}]$. 

\item[b)] Based on the pay-off division factor $\lambda^m_k$ and price $\mu^m$ broadcast by the MCO, the UUs estimate their pay-offs and decide to accept or reject the current coalition structure. If all the UUs are satisfied, go to Step 3). If at least one UU is not satisfied, it will proposed a new sub-band allocation which makes the current coalition structure invalid. Then go to Step 1-b). 
\end{itemize}
\item[Step 3)] Data transmission: 

After a stable coalition structure (i.e., sub-band and power allocation) is obtained, each UU starts data transmission with the optimal power calculated from (13). Note that the duration of data transmission should be less than the channel coherence time. %Furthermore, any join/leave action of the UU from the network will immediately interrupt the data transmission and the coalition structure should be re-formed. 
\end{itemize}

In each iteration, each of the ULSs will negotiate with $K-1$ other ULSs in a single sub-band. Considering there are $K$ ULSs and $M$ sub-bands, we can see that the time complexity is $\mathcal{O}((K-1)KM)$. 

we consider the communication overhead of the proposed protocol at the worst case. If we assume the size of the control packets in the proposed protocol is $v$ bits, then the overhead for channel gain estimation and neighborhood discovering is at most $[K + 2(K-1)]v$ bits. For the negotiation part, in each iteration there are at most $[(k-1)KM]v$ bits are sent. Recall that the coalition structure is proved to be $2^n$-finite, hence searching the core requires at most $2^K-1$ iterations. Therefore, in the worst case, the communication overhead will be $[(2^K-1)(k-1)KM + 3K -2 ]v$ bits. 

\subsection{Distributed Algorithm}
To reduce the number of iterations, we can use the similar way to that in \cite{xiao2012hierarchical} to drive the feasible region of the sub-band price $\mu_j$, which is given by $\mu_j \in [0, \overline{\mu}]$. Let $\overline{v}$ be the upper bound of $v_{S_k}$ and $\underline{h}$ be the lower bound of $|h_{jk}|^2$, then we have $\overline{\mu} = \frac{\overline{v}}{\underline{h}}$.

\begin{algorithm}
Step - 1)\textit{Sensing:} 
\begin{itemize}
\item[a)] The UUs, after receiving the prices of available sub-bands from the MCO, sequentially send a short training message to estimate their pay-off 
in all the sub-bands when the sub-bands are exclusively used by $S_k$.
\item[b)] Each $S_k$ broadcasts the sub-band combination $\bm{l}^*_{S_k}$ that maximizes its pay-off sum, 
\begin{equation}
\bm{l}^*_{S_k} = [l^{(1)}_{S_k}, l^{(2)}_{S_k},..., l^{(n)}_{S_k}].
\end{equation}
Let $\mathcal{R}^* = \{\bm{l}^*_{S_k}: S_k\in \{1, ..., K\}\}$.
\end{itemize}

Step - 2) \textit{Negotiation:}
\begin{itemize}
\item[a)]  All the active UUs need to negotiate with each other on each of the sub-bands in $\mathcal{R}^*$ to obtain the possible pay-off division factor $\lambda^m_{S_k}$. 
\item[b)]  After the negotiation process, $S_k$ solves problem (\ref{prob:pay_off_us}) based on the new set of $\lambda^m_{S_k}$, and obtains a new sub-band allocation to maximize its pay-off. 
Then $S_k$ updates and broadcasts its optimal sub-bands allocation.
Step 2) is repeated until no UU wants to change its occupied sub-bands.
\end{itemize}
\label{alg:alg1}
\caption{OCF Algorithm for Sub-band Allocation}
\end{algorithm}

\begin{algorithm}
\textit{Definitions:} At iteration $t$, let 
\begin{itemize}
\item[-] $\mu^m(t)$ be the pricing coefficient of sub-band $m$,
\end{itemize}

Step - 1) \textit{Initialization:}
\begin{itemize}
\item[-] Set $\mu^m \geq \overline{\mu}, \forall m\in \{1,2,...,M\}$.
\item[-] Set $\epsilon > 0$ to be a small positive constant.
\end{itemize}

Step - 2) \textit{Price Adjustment:}
\begin{itemize}
\item[a)] At iteration $t$, MCO updates and broadcasts $\bm{\mu}(t) = (1 - \epsilon)\bm{\mu}(t - 1)$.
\item[b)] Each $S_k$ senses the sub-bands and negotiates with other active UUs in the same sub-bands to determine the 
sub-band allocation $\bm{l}^{m*}(t)$ and power allocation $\bm{p}^{m*}(t)$.
\item[c)] All active UUs repeat Step 2-b) to update their optimal sub-bands, and the outcome is a coalition structure $\bm{P}^m(t)$. 
\item[d)] The MCO monitors the aggregated interference in each sub-band. If $N_j> \overline{Q}$, the price adjustment in sub-band $j$ stops.
If $N_j\leq \overline{Q}$, go to Step 2a).
\end{itemize}

Step - 3) \textit{Termination:} 

The algorithm ends with solution $\bm{\mu}^* = \bm{\mu}(t-1), \bm{P}^* = \bm{P}(t-1)$ in which the element 
$p^{m*}_{S_k}(\mu^{m*})$ is given by (\ref{eq:power_optt}).
\label{alg:alg2}
\caption{Distributed Interference Control Algorithm}
\end{algorithm}

Algorithms I %\ref{alg:alg1} 
and II
%\ref{alg:alg2} 
are proposed to find the SE of the hierarchical game.
For any given $\overline{Q}$, $\overline{p}$ pair and the channel gains, the algorithms achieve the SE which contains a stable overlapping coalition structure 
and an optimized power allocation for each UU. We have the following proposition about the SE of the game. 
\begin{myProp}
The price $\mu^m$ always converges to a non-negative value if a non-negative power allocation for a given $\overline{p}$ and $\overline{Q}$ pair exists. 
\label{prop:SG}
\end{myProp}

\begin{myProf}: 
See Appendix C. \qed
\end{myProf}

From propositions \ref{prop:OCFG} and \ref{prop:SG}, we conclude that, for any given $\bar{p}$ and $\overline{Q}$, the proposed algorithms will converge to the SE of the hierarchical game. The simulation results provided in section IV support this claim. 
\begin{myRmk}
The hierarchical game works as follows. At the beginning of iteration, the MCO broadcasts the  price $\bm{\mu}$ to all UUs in its coverage area. Each UU decides its optimal transmit power and sub-band based on the received pricing information sent 
by MCO. Once all UUs have made the decisions, MCO will adjust the price based on the interference before going to the next iteration.  
\end{myRmk}

The proposed algorithms can be implemented in a distributed manner. On the MCO side, it does not need to inquire any information from the UUs, e.g., the interfering link gain $h^m_{S_k}$ or corresponding transmit power $p^m_{S_k}$. It just measures the aggregated interference at its receiver in each channel, and adjusts the price accordingly. 
On the UUs side, with the channel price and the link gain information measured with in a coalition, they can easily derive the potential pay-off gained by joining  different coalitions. Therefore each of them can choose the best profited coalition combination to take part in. 

\begin{figure}
        \includegraphics[width=0.5\textwidth]{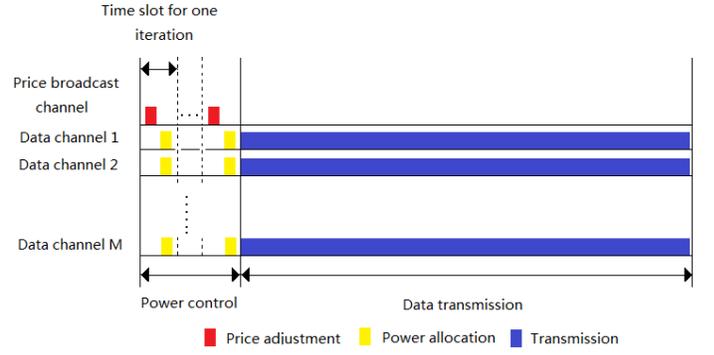}
        \caption{A time frame of proposed algorithm.}
    \label{A3:fig0}
\end{figure}

Considering the time overhead, for information exchange between the MCO and the UUs, there is a need for only one dedicated channel for the MCO to broadcast the interferences prices. The implementation is illustrated in Fig. \ref{A3:fig0}. 
A time frame for data transmission can be divided into two phases: the power control phase and the data transmission phase. In the power control phase, the time is divided into several time slots, which corresponds to an iteration in the proposed interference control algorithm. In each time slot, the MCO first measures the interference it is suffering, then adjusts the interference prices in each sub-band. Upon receiving the interference prices, the UUs re-allocate their power in each sub-band based on the prices and the measured mutual interference. After several iterations when the prices and power allocation are stable, each of the ULSs uses its power allocation in the last time slot to perform data transmission.  Suppose the price and power allocation will converge after $L$ time slots, each time slot duration is $\tau$, and the data transmission time is $t$, then the time overhead of the proposed algorithm should be 
$\frac{t}{K\tau+t}$. 

\section{Numerical Results}
In this section we investigate the performance of the proposed hierarchical game framework in the spectrum-sharing based femto-cell network. To better illustrate how to apply proposed algorithm adapts to various network environments, we consider the network system under different sets of interference and power constraints, as well as different numbers of UUs $K$ and available sub-bands $M$ combinations. The result shows that the proposed algorithm can automatically fit the constraints no matter which one dominates or both of them jointly apply.

\begin{figure}
    \centering
        \includegraphics[width=0.5\textwidth]{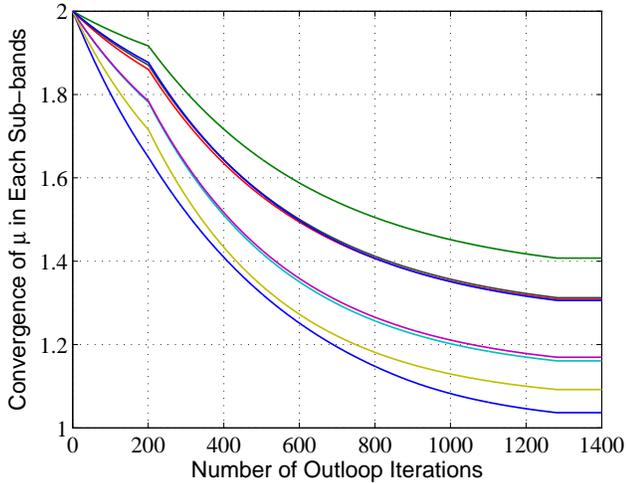}
        \caption{Convergence performance. $\overline{p} = 50$, $\overline{Q} = 2$. The curves illustrates the convergence of the interference prices in an $8$ sub-bands network. } 
    \label{fig:fig2}
\end{figure}

Fig. \ref{fig:fig2} illustrates the convergence of interference in a network with $8$ sub-bands, with $\overline{p} = 50$ and $\overline{Q}=2$. In Fig. \ref{fig:fig2}, the trend of the curves shows that the prices converge at around hundreds of iterations. Furthermore, it is noted that the prices in each sub-band converge at the similar speeds. This is because the prices of MCO directly control sub-band allocation and the power allocation of UUs.  %so the price in each sub-band will be affected by other oscillating prices, i.e., no one can converge solely. 
%However, an interesting observation is that some of the price curves oscillate and have sharp turning, because the coalition structure changes (which means a big change on the value of $\lambda^m_{S_k}$). Therefore the interference level has a sudden change in the sub-bands. 
Finally, the price charged to different sub-bands are independent with each other, which coincides with the definition in (\ref{eq:pay_off_mo}). 

\begin{figure}
    \centering
        \includegraphics[width=0.5\textwidth]{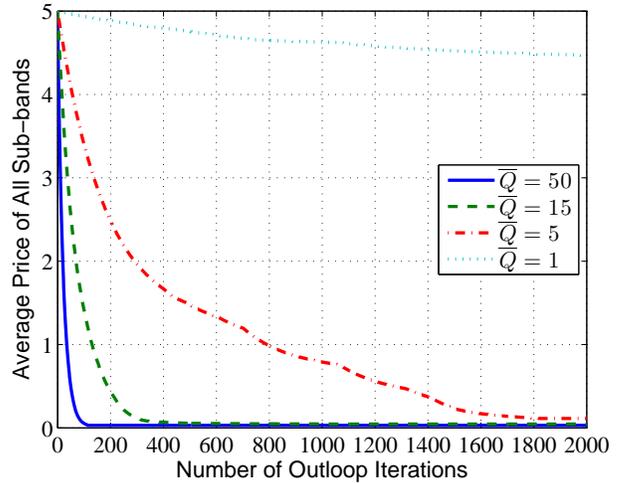}
        \caption{Convergence performance of the price under different  $\overline{Q}$, $\overline{p} = 100$. The curves shows the impact of $\overline{Q}$ on the convergence speed. } 
    \label{fig:fig2_add}
\end{figure}

In Fig. \ref{fig:fig2_add}, the convergence rate of average prices under different $\overline{Q}$ value is provided. An interesting observation is that, under the same power cap constraint, the convergence speed in the case of large $\overline{Q}$ is generally much faster than that in the case of small $\overline{Q}$. This phenomenon can be explained as follows: with the increase of $\overline{Q}$, each UU will allocate more power in each sub-band, hence under a fixed power cap constraint, each UU can access less sub-bands. Under tour setting, accessing less sub-bands is equivalent to join less coalitions. Hence, a large $\overline{Q}$ reduces the chances for UUs to join many coalition, which result in a reduced complexity for
coalition formation, and thus the time cost on forming a stable coalition structure can be  significantly reduced. 

\begin{figure}
\centering
%\subfigure[]{\includegraphics[width=0.4\textwidth]{Q_01_P_50_cut.eps}}\\
\includegraphics[width=0.5\textwidth]{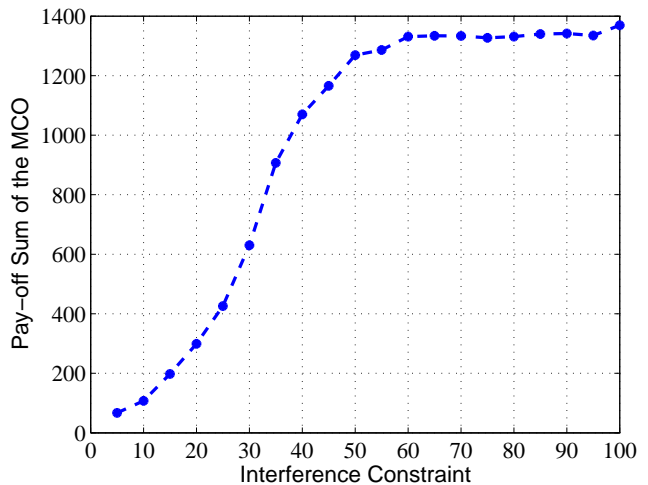}
  \caption[]{\label{fig:fig3}{} The impacts of varying the interference constraint: the pay-off sum increases with $\overline{Q}$.}
\end{figure}

\begin{figure}
\centering
%\subfigure[]{\includegraphics[width=0.4\textwidth]{Q_01_P_50_cut.eps}}\\
\includegraphics[width=0.5\textwidth]{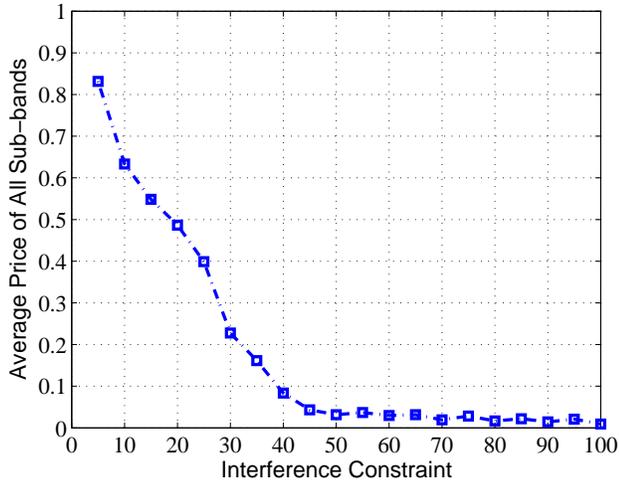}\\
  \caption[]{\label{fig:fig4}{} The impacts of interference constraint: the average price $\overline{\mu}$ decreases with $\overline{Q}$. }
\end{figure}

Figs. \ref{fig:fig3} to \ref{fig:fig4} show the convergence rate of the sub-band prices as well as the pay-offs of the MCO and UUs network. The tested network contains $64$ UUs and $128$ sub-bands, with $\overline{p} = 100$. 
Fig.\ref{fig:fig3} compares the pay-offs of the MCO versus the interference and power constraints. 
Assuming the channel coefficients are fixed, we increase one constraint while fixing the other one. It is observed that 
at the beginning of each time slot, the pay-offs increase with the constraint before they become steady. The reason for this is that 
initially the interference constraint is much tighter, which becomes the main limitation of the transmit power. However, when the interference constraint becomes 
larger, the transmit power is then jointly limited by both interference and power cap constraints. Finally when the interference constraint becomes very loose, the transmit power is limited by the 
power cap constraint so the system performance becomes stable. 

Fig. \ref{fig:fig4} illustrates the choice of interference limit $\overline{Q}$ against
the average price $\overline{\mu}$ over all sub-bands. The average price $\overline{\mu}$ generally reflects
the how much interference LUs can tolerate. It is observed that the price at $\overline{Q}=10$ is higher than that at $\overline{Q}=50$. 
This shows that the price decreases with the value of $\overline{Q}$. Generally speaking, the less the value of $\overline{Q}$, the rarer the resource is, so the price is accordingly larger. More specifically, it is obvious that the larger the $\overline{Q}$, the larger the possible transmit power of UU. If we look at the optimal power solution $p^{m*}_{S_k} = \left(\frac{1}{\mu^{m} h^m_{S_k}}-\frac{1}{\lambda^m_{S_k}} \right)^{+}$, we can see that $p^{m*}_{S_k}$ decreases with $\mu^m$, hence in sub-band $m$, a larger transmit power $p^{m*}_{S_k}$ results in a smaller interference price $\mu^m$. 

%being the amount of goods the MCO holds,  with the quantity of . If the amount of goods is
%small, it can be sold at a high price. Otherwise, if the
%vendor has a lot of goods to sell, he tends to sell them at a cheap price.

\begin{figure}
\centering
\includegraphics[width=0.5\textwidth]{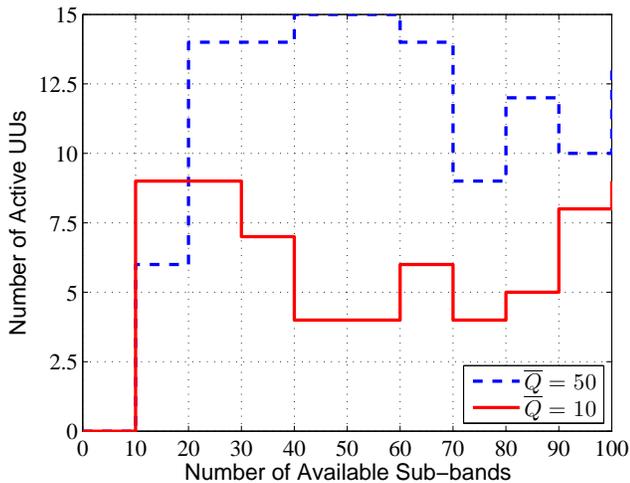}
  \caption[]{\label{fig:fig5}{} Comparison of $\overline{Q}=10$ and $\overline{Q}=50$. The number of active UUs versus the number of sub-bands. }
\end{figure}

\begin{figure}
\centering
\includegraphics[width=0.5\textwidth]{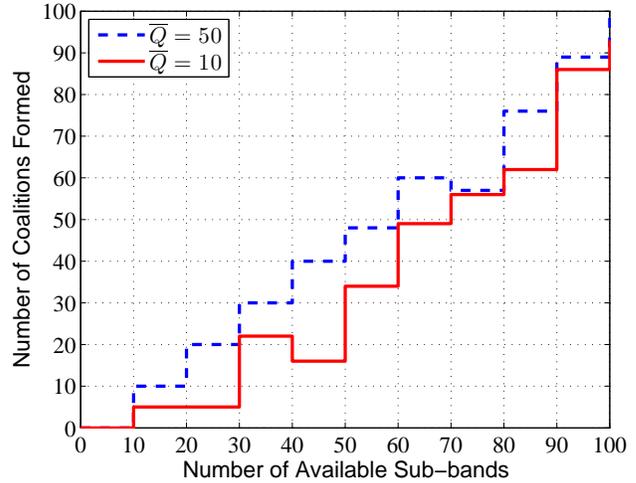}
  \caption[]{\label{fig:fig6}{} Comparison of $\overline{Q}=10$ and $\overline{Q}=50$.
	The number of coalitions versus the number of sub-bands. }
\end{figure}

\begin{figure}
\centering
\includegraphics[width=0.5\textwidth]{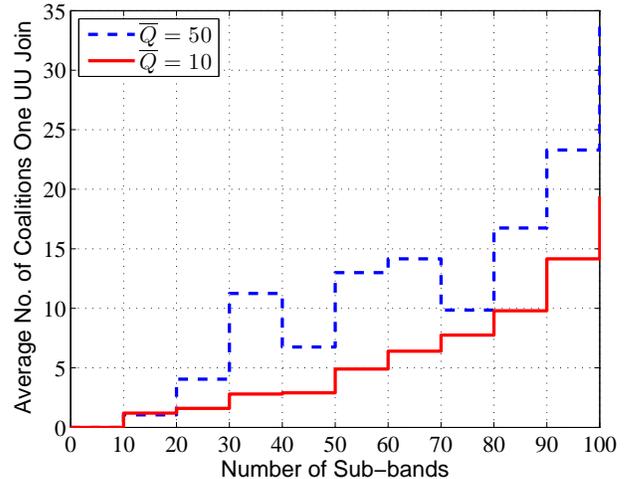}
  \caption[]{\label{fig:fig7}{} Comparison of $\overline{Q}=10$ and $\overline{Q}=50$. The average number of coalitions one UUs join against the number of available sub-bands.  }
\end{figure}

\begin{figure}
\includegraphics[width=0.5\textwidth]{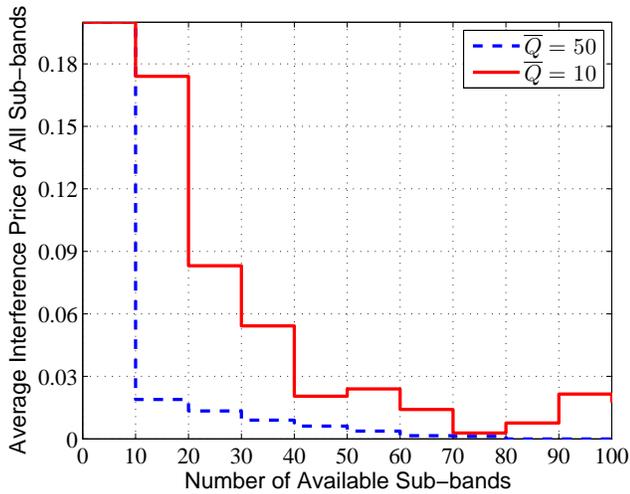}
  \caption[]{\label{fig:fig8}{} Comparison of $\overline{Q}=10$ and $\overline{Q}=50$. The average interference prices versus number of sub-bands. }
\end{figure}

Figs. \ref{fig:fig5} to \ref{fig:fig8} investigate the impact of the number of available sub-bands on the payoffs of UUs.
Fig. \ref{fig:fig5} and \ref{fig:fig6} shows the number of active UUs and the number of coalitions, under different numbers of sub-bands respectively. It is  seen that the number of active UUs is always lower than the total number of UUs. The reason is that if the channel gains of some UUs are highly correlated, the low payoff UUs will always be forced to leave the coalition. From Fig. \ref{fig:fig5}, it is observed that in general the larger $\overline{Q}$ the more active UUs, because larger $\overline{Q}$ enables more chances for the UU to transmit. Fig. \ref{fig:fig6} shows that the more available sub-bands the more coalitions formed, because when overlapping is enabled, the number of coalitions will be limited by the number of available sub-bands.

Fig. \ref{fig:fig7} and  \ref{fig:fig8} shows the average number of coalitions each UU joins and the average prices of sub-bands versus the number of sub-bands. Fig. \ref{fig:fig7} shows that the UU tends to join multiple coalitions when the number of available sub-bands increases, because in this case the players with lower pay-off in a crowded coalition may be better-off if joining a new coalition. Fig. \ref{fig:fig8} presents that the sub-band prices tend to decrease with the number of available sub-bands. When the UUs access multiple sub-bands, the aggregated interference in a single sub-band will be lower, which resulting lower sub-band prices. Another observation is that the price at $\overline{Q}=10$ is higher than that at $\overline{Q}=50$ because the tolerated interference is low when $\overline{Q}$ is small. Therefore the price is accordingly higher.

\begin{figure}
\centering
        \includegraphics[width=0.50\textwidth]{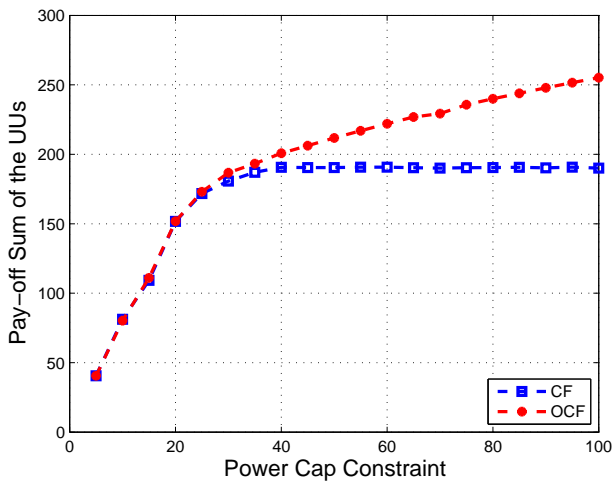}
  \caption{The comparison between CF and OCF schemes.}
\label{C4:fig7}
\end{figure}

Figure \ref{C4:fig7} compares the proposed OCF algorithm with the traditional coalition formation setting without overlapping. It is illustrated directly in the figure that the improvement of data rate by enabling overlapping. When the power available for transmit goes high, the UUs in OCF scheme are benefited by exploring more chances to transmit on multiple sub-bands while in the CF schemes each of the UU can only access a single sub-band.

%multi-user-mulit-band access problem and the power allocation issues based on a cooperative game model 
\section{Conclusion}
The sub-band allocation and the power control issues in the carrier-aggregation-enabled HetNet are studied in this paper. 
We have developed a hierarchical game framework to jointly solve the power and sub-band allocation problems under the constraints of the transmit power and 
maximum tolerable interference level. 
We established a Stackelberg game for MCO to regulate the transmit power of the UUs so as to give sufficient protection to the LUs while optimizing the pay-off of the UUs. 
We also apply OCF-game to analyze the behavior of the UUs that can self-organize into overlapping coalitions. 
We have proposed a simple two-layer algorithm to let the UUs iteratively search for the 
optimal coalition structure and the power allocations under different prices imposed by the macro-cell.
It has been proved that the proposed algorithm can always 
converges to the SE of hierarchical game and at the same time the resulting transmit power and the sub-band allocations are stable and no players can further improve their payoff by unilaterally deviate from it 
by acting alone. 
Furthermore, by allowing the overlapping in the coalition formation among UUs,, we have addressed the problem of sub-band and power allocation problem under two dimension constraints. The proposed framework can also be extended into more general network setting with multiple BSs to cooperatively share their sub-bands or the downlink communication that multiple LUs need to be protected.  

\begin{appendices}
      \section{Proof of Proposition $1$}
      %\begin{myProff}
Suppose a partial coalition $\bm{p}^{m*} = \{p^{m*}_{S_k}: k=1, 2, ...K\}$ is formed on sub-band $m$, in which the positive power $p^{m*}_{S_k}$ is given by 
(\ref{eq:power_optt}), i.e., 
\begin{align}
\bm{p}^{m*} = \arg \max \limits_{\bm{p}^{m}} \pi(\bm{p}^{m}). 
\label{eq:partial_opt}
\end{align}
We define the support of $\bm{p}^{m*}$ as,
\begin{align}
\mbox{supp}(\bm{p}^{m*}) = \{S_k:p^{m*}_{S_k} >0, k=1, 2, ...K\}^m, 
\end{align}
which defines a coalition of UUs regardless the resource distribution. 
Hence, for any other partial coalition $\bm{p}^{m'}$ with the support $\mbox{supp}(\bm{p}^{m*})$, we have
\begin{align}
\pi(\bm{p}^{m*}) \geq \pi(\bm{p}^{m'}),
\end{align}
i.e., the partial coalition $\bm{p}^{m*}$ \textit{blocks} all other partial coalitions formed on sub-band $m$ which involves with $\mbox{supp}(\bm{p}^{m*})$. 

Therefore, we can say that the partial coalition $\bm{p}^{m*}$ in our proposed game is one-to-one correspondent to the coalition $\{S_k:p^{m*}_{S_k} >0, k=1, 2, ...K\}^m$ formed 
on sub-band $m$. Since $\{S_k\}^m \subseteq \mathcal{K}$, i.e., $\{S_k\}^m$ is a subset of  $\mathcal{K}$, the number of all possible partial coalitions equals to the number of subset of $\mathcal{K}$, which is given by, 
\begin{align}
\sum_{n=1}^{K}{\binom{K}{n}} = 2^K - 1. 
\end{align}
This concludes the proof.\qed
%\cite{telatar1999capacity}
%\end{myProff}
      \section{Proof of Proposition $2$}
%\begin{myProff}:
\begin{itemize}
\item[1)]
Continuous. 
The value function in (\ref{eq:value_fun}) is the difference between a log function and a linear function, 
which is obviously continuous.
\item[2)]
Monotone.
The interference power constraint in (\ref{co:co1}) limits the total transmit power allocated in sub-band $m$ indirectly by pricing in the 
Stackelberg game. Hence the power allocated by $S_k$ in sub-band $m$ is bounded by $p^{m*}_{S_k}$. 
Since the pay-off function, $\pi(p^m_{S_k})$, of $S_k$  is concave, then for any 
$\pi(p^{m'}_{S_k}) \in [0, p^{m*}_{S_k}]$ we have $\pi(p^{m'}_{S_k}) \leq \pi(p^{m*}_{S_k})$. Therefore for any $\bm{p}^{m'}$ and $\bm{p}^{m*}$, such that 
 $p^{m'}_{S_k} \leq p^{m*}_{S_k}$, we have $\bm{v}(\bm{p}^{m'}) \leq \bm{v}(\bm{p}^{m*})$, i.e., the value function is monotone. 
\item[3)]
Bounded. 
According to the proof in 2), the value function is bounded by $\bm{v}(\bm{p}^{m*})$, where $\bm{p}^{m*} = (p^m_{S_1}, p^m_{S_2}, ..., p^m_{S_K})$ satisfies 
$\sum_{k=1}^K{h^m_{S_k}p^m_{S_k}} = \overline{Q}$. 
\item[4)]
U-finite.
The proof can be referred to proposition \ref{prop:U_finite}. 
\item[5)]
The inequality. The equality of (\ref{eq:eq_16}) is always taken in the proposed game since the value function is the summation of individual pay-off of the member players. 
\end{itemize}
\qed
%\end{myProff}
      \section{Proof of Proposition $3$}
%\begin{myProff}: 
In previous section we proved that finding optimal pricing using $\mu^{m*} = \arg \max \limits_{\mu^m} \pi_{MCO}(\bm{p}^*, \mu^m)$ 
is equivalent to solving $\sum_{k=1}^{K}{\left(\frac{1}{\mu^{m*} h^m_{S_k}}-\frac{1}{\lambda^m_{S_k}} \right)^{+}}h^m_{S_k} = \overline{Q}$. 
Hence the pay-off maximizing for MCO can be achieved by choosing the optimal price to control the interference approaching $\overline{Q}$. 
In other words, the only two cases that the MCO will stop further increasing or decreasing prices are, 
1) $\sum_{k=1}^{K}{p^m_{S_k}h^m_{S_k}} \leq \overline{Q}$, and 2) $\mu^m=0$. 
In other words, the price $\mu^m$ can always converge to a fixed price. \qed
%\end{myProff}

\end{appendices}


\end{document}